\documentclass[conference]{IEEEtran}
\IEEEoverridecommandlockouts
\makeatother
\usepackage{cite}
\usepackage{amsmath,amssymb,amsfonts}
\usepackage{algorithmic}
\usepackage{graphicx}
\usepackage{textcomp}
\usepackage{booktabs}
\usepackage{xcolor}
\usepackage{float}
\def\BibTeX{{\rm B\kern-.05em{\sc i\kern-.025em b}\kern-.08em
    T\kern-.1667em\lower.7ex\hbox{E}\kern-.125emX}}
\begin{document}

\title{Advanced Payment Security System:
\\XGBoost, LightGBM and SMOTE Integrated 
}

\author{
\IEEEauthorblockN{1\textsuperscript{st} Qi Zheng}
\IEEEauthorblockA{
    \textit{Northeastern University} \\
    Boston, MA, 02115, USA\\
    Email: zheng.qi2@northeastern.edu}
\and
\IEEEauthorblockN{1\textsuperscript{st*} Chang Yu}
\IEEEauthorblockA{
    \textit{Northeastern University} \\
    Boston, MA, 02115, USA\\
    Email: chang.yu@northeastern.edu}
\and
\IEEEauthorblockN{2\textsuperscript{nd} Jin Cao}
\IEEEauthorblockA{
    \textit{Johns Hopkins University}\\
    Baltimore, MD, 21218, USA \\
    Email: caojinscholar@gmail.com}
\and
\IEEEauthorblockN{2\textsuperscript{nd} Yongshun Xu}
\IEEEauthorblockA{
    \textit{University of Massachusetts Lowell} \\
    Lowell, MA, 01850, USA\\
    Email: Yongshun\_Xu@student.uml.edu}
\and
\IEEEauthorblockN{3\textsuperscript{rd} Qianwen Xing}
\IEEEauthorblockA{
    \textit{University of Chicago} \\
    Chicago, IL, 60637, USA \\
    Email: xqw3669@gmail.com}  
\and
\IEEEauthorblockN{3\textsuperscript{rd} Yixin Jin }
\IEEEauthorblockA{
\textit{University of Michigan, Ann Arbor}\\
Ann Arbor, MI 48109, USA \\
Email: jyx0621@gmail.com}
}

\maketitle

\begin{abstract}
  With the rise of various online and mobile payment systems, transaction fraud has become a significant threat to financial security. This study explores the application of advanced machine learning models, specifically based on XGBoost and LightGBM, for developing a more accurate and robust Payment Security Protection Model. To enhance data reliability, we meticulously processed the data sources and applied SMOTE (Synthetic Minority Over-sampling Technique) to address class imbalance and improve data representation. By selecting highly correlated features, we aimed to strengthen the training process and boost model performance. We conducted thorough performance evaluations of our proposed models, comparing them against traditional methods including Random Forest, Neural Network, and Logistic Regression. Using metrics such as Precision, Recall, and F1 Score, we rigorously assessed their effectiveness. Our detailed analyses and comparisons reveal that the combination of SMOTE with XGBoost and LightGBM offers a highly efficient and powerful mechanism for payment security protection. Moreover, the integration of XGBoost and LightGBM in a Local Ensemble model further demonstrated outstanding performance. After incorporating SMOTE, the new combined model achieved a significant improvement of nearly 6\% over traditional models and around 5\% over its sub-models, showcasing remarkable results.

\end{abstract}


\section{Introduction}
Across the globe, network payment platforms are emerging as a burgeoning industry. With the widespread adoption of online payments\cite{ELHEDHLI2017165}, these platforms have brought about profound and extensive changes to society, fundamentally altering our lifestyles. Network payments involve issues related to fund security and are closely linked to societal stability. The problems inherent in online payments have long constrained the development of e-commerce in the modern financial sector. Various phenomena such as transaction fraud and payment fraud frequently occur, causing significant distress for both bankers and users, and represent one of the primary obstacles to the development of network payments.

To address these issues, predecessors have summarized and explored various powerful and reliable classification algorithms, including Neural Networks, Logistic Regression, and SVM. However, these algorithms still face challenges such as complex hyperparameter tuning, handling multi-class problems, long training times, and unsuitability for large-scale datasets. The newly introduced LightGBM and XGBoost have made significant improvements in large-scale data processing, nonlinear handling, training speed, and performance tuning. Nonetheless, they still have drawbacks in handling nonlinear relationships and susceptibility to outliers. Combining their strengths for comprehensive optimization remains a highly challenging task.

To build on the foundation of these models, we plan to combine various technologies and models to explore more powerful and effective solutions. In this article, we will focus on some of the most commonly used models in banking today . After processing fraud data features, we will compare XGBoost, LightGBM, CatBoost, and traditional models. We will then attempt to combine these models and further optimize them using SMOTE. Through detailed experimental comparisons of parameters such as F1 score, precision, recall, and AUC, we aim to develop a new model.

The new model is a composite model based on a stacking architecture, using LightGBM and XGBoost as meta-models and a Neural Network as the secondary classifier. This approach cleverly combines these three different types of models. As a result, we obtain a new model with better overall performance, higher accuracy, better fault tolerance, and broader applicability.

Next, in Section II, we will introduce the related work behind this article and the sources of our data. Section III will detail the technologies we used, including LightGBM, XGBoost, and the new ensemble model built on top of them, and provide a thorough explanation of our data processing and feature processing methods. This section will give an overview of the specific techniques and frameworks we employed. In Section IV, we will describe our experimental procedures in detail, including the various data processing methods used in the experiments, the steps of the processing, and the related results. Finally, in Section V, we will summarize the experimental results, outline our proposed directions for application, and lay the groundwork for future research.

\section{Related work}
For this study, we thoroughly reviewed and compared various previous research models, summarizing their advantages and potential performance bottlenecks. Based on the distinct characteristics of these models, we integrated them in different ways, leading to the results of our current research.

Sayjadah has implemented machine learning models to build a robust payment security system to prevent fraud in transactions. He pioneered the use of logistic regression, decision trees, and random forests, achieving an accuracy rate of 82\% in risk prediction. While his work laid a solid foundation for our research direction, it also has limitations such as poor hyperparameter performance, unsuitability for large-scale data processing, and inadequate performance in handling multi-class classification problems.

Chen and Guestrin (2016) provided a detailed explanation of XGBoost's principles and training characteristics. Their research has significantly influenced our work, as we adopted many of their methodologies to enhance our results. XGBoost's efficiency and performance improvements make it an essential component of our model ensemble. LightGBM (Light Gradient Boosting Machine), developed by Ke et al. (2017) , is another powerful gradient boosting framework designed for speed and efficiency. We also integrated LightGBM into our study due to its ability to handle large-scale data with high efficiency, making it a crucial model in our experiments.Although both models are excellent, they also have their own drawbacks and issues, such as overfitting, sensitivity to outliers, and training time.

For our ensemble model, we reviewed various models that matched our requirements. Ultimately, inspired by Seewald, we decided to build our new classifier using the stacking approach. Seewald's work on improving the effectiveness and efficiency of stacking provided a strong foundation for our ensemble method.

The article "A Survey on Applications of Artificial Intelligence for Smart Manufacturing: Real-Time Production Quality Improvement and System Optimization" offers significant insights that can be directly applied to the development of advanced payment protection systems \cite{survey2020}. This survey highlights how artificial intelligence (AI) techniques , such as machine learning, deep learning, and reinforcement learning, are revolutionizing manufacturing processes by improving quality and optimizing system performance.The aforementioned article serves as an important reference for us in model construction and processing.

Outlier detection and handling is an indispensable component of data preprocessing . By undertaking this procedure, we aimed to ensure the quality and integrity of the data, enhance the performance of subsequent machine learning models, uncover valuable insights embedded within the data\cite{peng2024maxk}, mitigate the business impact of anomalies, and satisfy relevant regulatory and compliance requirements.

\section{Methology}
In this chapter, we have detailed the foundational architecture and principles of our Payment Security System. Next, we will introduce the specific principles of the models used in our experiments. Following that, we will discuss the data analysis and optimizations we performed to achieve better experimental results.

\subsection{Payment Security System Architecture}\label{AA}
As shown in Figure 1, our system is based on accumulated historical transaction data. Depending on different applications and previous experimental results, techniques such as Logistic Regression and Neural Networks have already been widely applied in similar system models. These models perform an initial classification of new transaction requests\cite{jin2024apeer}, approving legitimate payments and forwarding potential fraud transactions to the Fraud Protection Service for further processing. Therefore, the accuracy of the models is crucial to the system's performance.

In this research, we will use a series of techniques including XGBoost, LightGBM, and CatBoost to process the data and perform comparisons to identify the best-performing model\cite{jin2024learning}. Subsequently, we will employ the stacking ensemble technique to construct a new, more powerful ensemble model. This model, combined with SMOTE (Synthetic Minority Over-sampling Technique), will be compared against a series of classic models for research and validation.

\begin{figure}[h]
    \centering
    \includegraphics[width=1\linewidth]{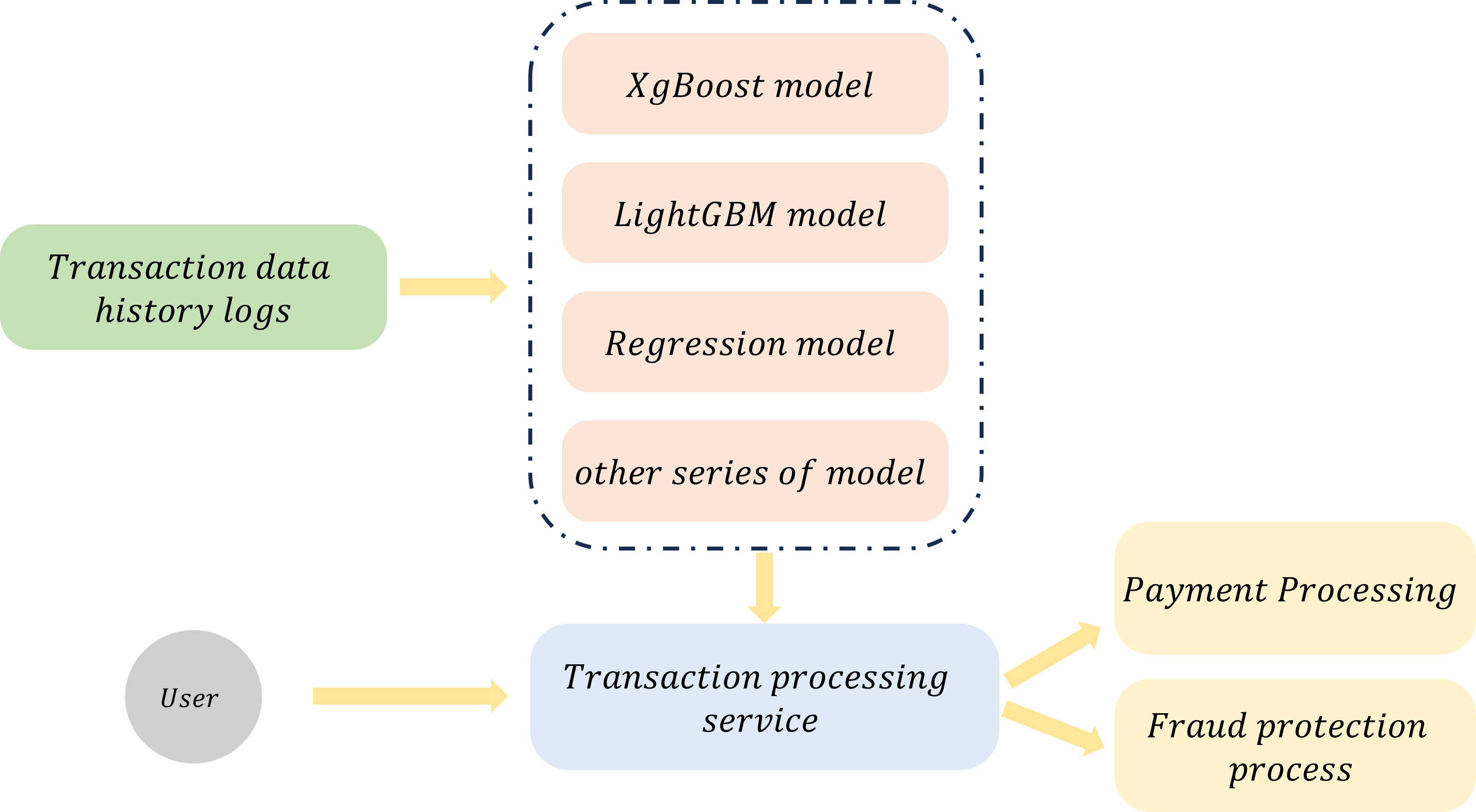}
    \caption{Ensemble model structure.}
    \label{fig:enter-label}
\end{figure}

\subsection{Data Processing}
In our data processing phase, the primary objectives were to clean missing values and outliers, as well as to select the most relevant features for training. This preprocessing is crucial to enhance the effectiveness of our model training. The steps involved in our data processing are outlined below:
\subsubsection{Noise Removal}
Noise in the data can obscure the proper signal and lead to poor model performance. We applied noise removal techniques, such as smoothing methods\cite{li2021online}, filtering, and transformations, to eliminate irrelevant or random variations in the data. This helps to enhance the signal-to-noise ratio, making the underlying patterns more discernible and improving the accuracy of our models.The result is shown in Fig2.

\begin{figure}[h]
    \centering
    \includegraphics[width=0.9\linewidth]{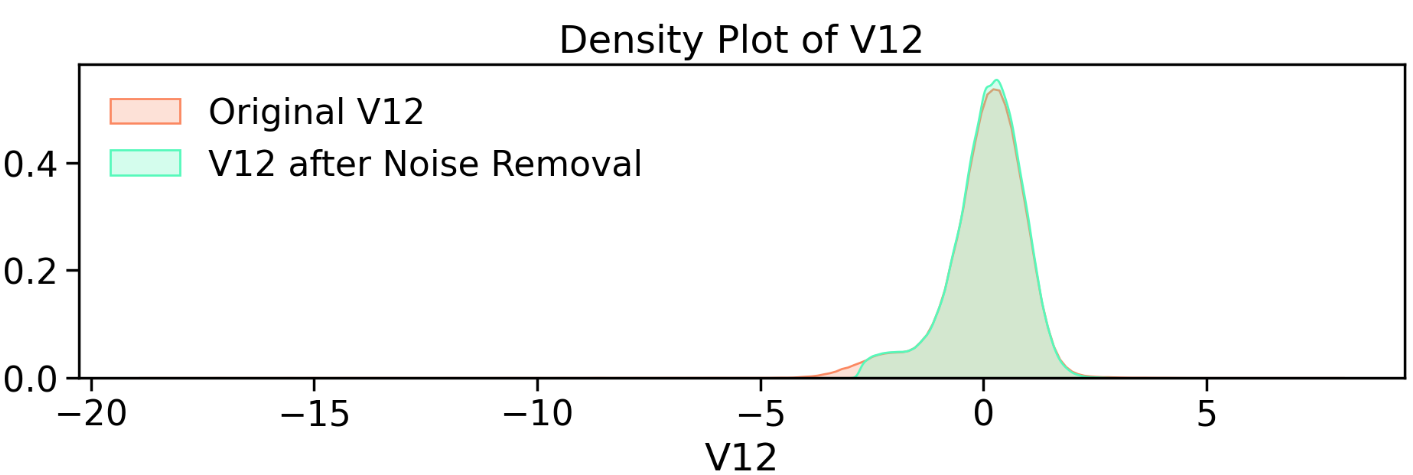}
    \caption{Density comparison after noise removal.}
    \label{fig:enter-label}
\end{figure}

We use the Z-score method here. The Z-score standardizes data by calculating the distance of each data point from the mean (measured in standard deviations). Specifically, the Z-score represents how much a data point deviates from the mean. Typically, data points with Z-scores exceeding a certain threshold are considered outliers or noise and can thus be removed to improve the quality and reliability of the data.

\subsubsection{Anomaly Detection and Outlier Removal}
Anomaly detection and outlier removal are crucial steps in data processing that can significantly enhance model performance, prevent overfitting, improve data consistency, and increase training effectiveness. In this analysis, we employed the Interquartile Range (IQR) method for anomaly detection and similarly used the IQR technique for outlier removal\cite{jiang2022weakly}. Below are graphical examples illustrating the results after outlier removal.

Here we use the IQR method. The IQR (Interquartile Range) method is used to detect and remove outliers in the data. The specific steps include calculating the 25th and 75th percentiles (Q1 and Q3) of the feature values, and then determining the upper and lower bounds for outliers based on the IQR (Q3 - Q1), typically 1.5 times the IQR. Any values exceeding these bounds are considered outliers and are removed from the data to improve the quality and reliability of the model. The specific results are shown in the following figure.
\begin{figure}[H]
    \centering
    \includegraphics[width=0.5\linewidth]{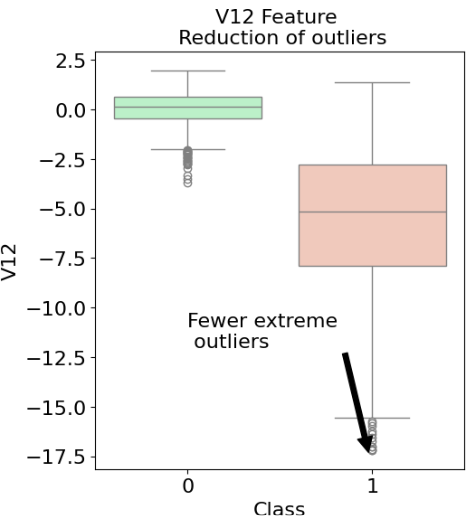}
    \caption{Feature distribution comparison after outlier removal.}
    \label{fig:enter-label}
\end{figure}

\subsubsection{Dimensionality reduction}
In this data processing task, we employed the T-SNE technique to perform dimensionality reduction on high-dimensional data, achieving remarkable results. T-SNE uses non-linear mapping to project data points from a high-dimensional space to a low-dimensional space while preserving the similarity structure between data points as much as possible. This allows us to visually observe the intrinsic patterns and clustering structure of the data in a two-dimensional or three-dimensional space, thereby better understanding the characteristics and distribution of the data. Dimensionality reduction brings multiple benefits. it significantly reduces computational complexity, eliminates redundant information, solve the data sparse problem.

The working principle of T-SNE can be summarized in the following steps: First, it calculates the similarity between data points in the high-dimensional space, typically using Euclidean distance or other distance metrics. Then, based on the similarity, it transforms the data points into a conditional probability distribution. In the low-dimensional space, T-SNE initializes the positions of the data points and calculates the similarity between data points in the low-dimensional space\cite{yang-24-data-aug}, also represented by a conditional probability distribution. Next, by minimizing the difference between the two conditional probability distributions in the high-dimensional and low-dimensional spaces (usually using KL divergence), T-SNE iteratively optimizes the positions of the data points in the low-dimensional space, making similar data points in the high-dimensional space also closer in the low-dimensional space. Finally, the iterative optimization continues until convergence is reached, resulting in the final low-dimensional representation.
\begin{figure}
    \centering
    \includegraphics[width=0.65\linewidth]{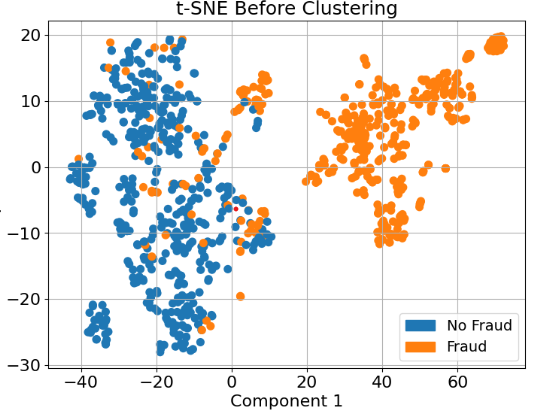}
    \caption{Data distribution before clustering.}
    \label{fig:enter-label}
\end{figure}

\begin{figure}
    \centering
    \includegraphics[width=0.65\linewidth]{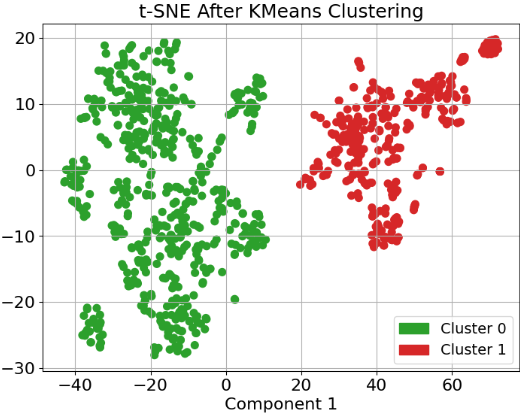}
    \caption{Data distribution after clustering.}
    \label{fig:enter-label}
\end{figure}

\subsubsection{Synthetic Minority Over-sampling Technique (SMOTE)}
When dealing with imbalanced datasets, classifiers often tend to favor the majority class, overlooking the minority class samples. This bias significantly reduces the model's performance on the minority class, which can be critical in scenarios where minority class detection is essential. To address this issue, this study employs the Synthetic Minority Over-sampling Technique (SMOTE) to balance the dataset.

SMOTE generates synthetic samples by interpolating between minority class samples in the feature space. Specifically, SMOTE selects a random sample \(x_i\) from the minority class and identifies its \(k\)-nearest neighbors. It then randomly selects one of these neighbors \(x_{i,k}\) and generates a new synthetic sample \(x_{\text{new}}\) along the line segment between \(x_i\) and \(x_{i,k}\) using the formula:
\[
x_{\text{new}} = x_i + \lambda \cdot (x_{i,k} - x_i)
\]
where \(\lambda\) is a random number in the range of 0 and 1\cite{jordan2018learning}. This method ensures that the generated samples reflect the characteristics of the minority class more accurately, thereby enhancing the classifier's ability to recognize minority class instances. 

By incorporating SMOTE, we effectively balanced the class distribution in the training phase, significantly improving the classifier's performance on the minority class. This approach not only enhances the overall accuracy of the model but also improves recall and precision for the minority class, validating the effectiveness of SMOTE in handling imbalanced datasets.

\subsection{Models}
\subsubsection{LightGBM model}
LightGBM (Light Gradient Boosting Machine) is a highly efficient and distributed gradient boosting framework developed by Microsoft\cite{ke2017lightgbm}. LightGBM introduces several innovative techniques, including Gradient-based One-Side Sampling, Exclusive Feature Bundling, and histogram-based decision tree algorithms. These algorithms, combined with optimizations for parallel and distributed learning, make LightGBM particularly suitable for large-scale learning tasks. Based on these advantages, we have incorporated this model in our experiments.

\subsubsection{ XGBoost Model}
XGBoost, developed by Tianqi Chen and his team, is a high-performance, scalable implementation of the gradient boosting algorithm. It is renowned for its exceptional performance across various tasks, featuring regularization (L1 and L2) to prevent overfitting, efficient sparse data handling, support for parallel and distributed computing, and advanced optimization methods like sparsity-aware split finding and histogram-based decision tree construction. These advantages make XGBoost ideal for large-scale learning tasks, and thus we incorporated it into our experiments\cite{chen2016xgboost}.

\subsubsection{Esmble Model}
To obtain the best factors for data interaction among several models, we used an Ensemble model in our design. After multiple attempts, we constructed an Ensemble Model using XGBoost and LightGBM as base classifiers, with a neural network as the final classifier.

In the model design, we used XGBoost and LightGBM as elements of the base classifier layer. Assume we have a training set \( D = \{ (x_i, y_i) \}_{i=1}^N \), where \( x_i \) is the important vector of the \( i \)-th sample, and \( y_i \) is the correspondng label. We train \( M \) base classifiers \( \{ h_j \}_{j=1}^M \), where each classifier \( h_j \) is trained using the entire training set. For the base classifiers \( h_j \) constructed by LightGBM and XGBoost, we can obtain the predicted values for all samples in the training set, \( y_{ij} = h_j(x_i) \).

We combine the predictions of all base classifiers into a new feature vector. For each sample \( x_i \), the new feature vector \( z_i \) is represented as:
\[ z_i = [y_{i1}, y_{i2}, \ldots, y_{iM}] \]

Our model design incorporates a neural network-based meta-classifier, \( h_{\text{meta}} \), which is trained using the new feature vectors \( \{ z_i \}_{i=1}^N \) and the original labels \( \{ y_i \}_{i=1}^N \). The primary goal of \( h_{\text{meta}} \) is to learn to combine the predictions of the base classifiers, thereby enhancing the overall prediction accuracy.

The training process of the meta-classifier can be expressed as:
\[ h_{\text{meta}} = \arg\min_h \sum_{i=1}^N L(y_i, h(z_i)) \]
where \( L \) is the loss function, such as Mean Squared Error (MSE) or Cross-Entropy Loss.

While building this model, for the sample \( x_{\text{new}} \), we first use the base classifiers LightGBM and XGBoost to obtain their predictions and then use these predictions as inputs to the neural network-based meta-classifier to get the final prediction:
\[ y_{\text{new}} = h_{\text{meta}}([y_{\text{new},1}, y_{\text{new},2}, \ldots, y_{\text{new},M}]) \]
where \( y_{\text{new},j} = h_j(x_{\text{new}}) \) is the prediction of the \( j \)-th base classifier for the new sample.

\section{Evaluaion}
In the following sections, we will utilize various metrics to conduct comprehensive comparisons and experiments, involving all the models mentioned in the methods section. The results of these experiments, along with detailed analyses and interpretations, will be presented and thoroughly discussed in the subsequent chapters.
\subsection{Evaluation Metrics}
In this experiment, we used F1 Score, Precision, Recall, and AUC Score to evaluate our model performance.

Precision measures the true positive instances among those predicted as positive, reflecting the model's accuracy.

Recall indicates the proportion of actual positive instances correctly identified, showing the model's ability to capture all relevant instances.

F1 Score is the harmonic mean of Precision and Recall\cite{sharma2023neural}, providing a balanced measure that considers both accuracy and coverage.

AUC Score evaluates the model's ability to distinguish between classes and is particularly useful for imbalanced datasets. A higher AUC indicates better performance.

Using these metrics, we developed a robust framework to meticulously assess and select the most effective models, ensuring reliability and insight in handling complex data.

By leveraging these metrics, we have developed a robust and efficient framework to conduct precise and objective analyses of our data. This comprehensive evaluation process enables us to thoroughly assess various aspects of the models' performance. Consequently, we can identify and select the most effective and powerful models, ensuring that our approach remains both reliable and insightful in addressing the complexities of the data.
\subsection{Experiement Result}
In this section, we conduct a detailed comparison of various models used in a Fraud Protection System. To ensure the reliability of our model comparison, we not only employed traditional models such as Logistic Regression and Decision Tree but also incorporated some of the most popular high-performance models available today. This includes deep learning-based models like TabNet, CatBoost, and standalone models such as LightGBM and XGBoost. This comprehensive comparison aims to evaluate the detailed performance of each model while also assessing their practical utility in modern-day scenarios.

\begin{table}[h]
\centering
\caption{Performance comparison of various classifiers.}
\label{table:comparison}
\begin{tabular}{|l|c|c|c|c|}
\hline
\textbf{Classifier} & \textbf{Precision} & \textbf{Recall} & \textbf{F1 Score} & \textbf{AUC} \\ \hline
Logistic Regression & 0.94 & 0.90 & 0.92 & 0.97 \\ \hline
K-Nearest Neighbors & 0.96 & 0.87 & 0.91 & 0.96 \\ \hline
Support Vector Classifier & 0.93 & 0.93 & 0.93 & 0.93 \\ \hline
Decision Tree Classifier & 0.89 & 0.90 & 0.89 & 0.90 \\ \hline
XGBoost & 0.94 & 0.90 & 0.92 & 0.97 \\ \hline
LightGBM & 0.95 & 0.89 & 0.92 & 0.97 \\ \hline
CatBoost & 0.96 & 0.89 & 0.92 & 0.97 \\ \hline
TabNet & 0.83 & 0.71 & 0.76 & 0.84 \\ \hline
Neural Network & 0.93 & 0.91 & 0.92 & 0.97 \\ \hline
Ensemble Model & 0.97 & 0.86 & 0.91 & 0.97 \\ \hline
\end{tabular}
\end{table}

From the data comparison, we can see that the Ensemble Model performs exceptionally well on most metrics, particularly in terms of precision (0.97) and AUC (0.97), demonstrating its advantage in overall performance. Although the recall is slightly lower (0.86), it remains within an acceptable range. The Ensemble Model effectively reduces the bias and variance of individual models by combining the prediction results of multiple base models, thus enhancing the robustness and generalization ability of the model. The Ensemble Model is suitable for various data types and tasks, whether classification or regression problems. It can optimize performance by adjusting the base models and ensemble strategies, offering significant flexibility. By combining the strengths of multiple models, the Ensemble Model shows outstanding overall performance and robustness, providing high precision and AUC in most tasks. Although the recall is slightly lower, the overall performance remains excellent, making it suitable for various data types and application scenarios.

After applying SMOTE processing to the data, we further trained the models mentioned above. We can clearly see that after SMOTE processing, all the models were optimized to varying degrees.

\begin{table}[h]
\centering
\caption{Performance comparison of various classifiers after SMOTE processing.}
\label{table:smote_comparison}
\begin{tabular}{|l|c|c|c|c|}
\hline
\textbf{Classifier} & \textbf{Precision} & \textbf{Recall} & \textbf{F1 Score} & \textbf{AUC} \\ \hline
Logistic Regression & 0.96 & 0.92 & 0.94 & 0.98 \\ \hline
K-Nearest Neighbors & 0.96 & 0.88 & 0.92 & 0.97 \\ \hline
Support Vector Classifier & 0.99 & 0.87 & 0.92 & 0.98 \\ \hline
Decision Tree Classifier & 0.89 & 0.92 & 0.91 & 0.92 \\ \hline
XGBoost & 0.96 & 0.92 & 0.94 & 0.98 \\ \hline
LightGBM & 0.97 & 0.91 & 0.94 & 0.98 \\ \hline
CatBoost & 0.97 & 0.91 & 0.94 & 0.98 \\ \hline
TabNet & 0.93 & 0.94 & 0.93 & 0.96 \\ \hline
Neural Network & 0.96 & 0.93 & 0.94 & 0.98 \\ \hline
Ensemble Model & 0.98 & 0.90 & 0.94 & 0.99 \\ \hline
\end{tabular}
\end{table}

By comparing the above parameters and images, we can see that the precision of the Ensemble Model is 0.98, the highest among all models, tied with the Support Vector Classifier. This means that among all samples predicted as positive, the Ensemble Model has the highest correct prediction rate and the lowest false positive rate. Additionally, the AUC score of the Ensemble Model is 0.99, the highest among all models. The higher the AUC score, the stronger the model's performance to identify positive and negative samples. The Ensemble Model performs best in this regard, indicating its superior overall classification ability.

Even in terms of Recall and F1 Score, the recall rate of the Ensemble Model is 0.90, which, although not the highest, is not significantly different from other models. The slightly lower recall rate could be because the Ensemble Model emphasizes precision more, but it still maintains a high overall level. The F1 Score balances Precision and Recall, with the Ensemble Model achieving an F1 Score of 0.94, consistent with Logistic Regression, XGBoost, LightGBM, CatBoost, and Neural Network. This indicates that the Ensemble Model performs excellently in balancing precision and recall.

As an ensemble model that combines multiple base models, the Ensemble Model integrates the prediction results of several base models (such as Logistic Regression, XGBoost, Neural Network, etc.), fully leveraging the strengths of each model and compensating for the shortcomings of single models. Because it aggregates the results of multiple models, the Ensemble Model has better resistance to overfitting and underfitting issues of individual models, making the overall model more stable. By combining multiple models, the Ensemble Model can better capture the diversity of data distributions, improving generalization to new data.

In summary, the Ensemble model combined with SMOTE demonstrates outstanding performance across various metrics, particularly excelling in precision and AUC value, highlighting its superiority in practical applications.

\section*{Conclusion}

In today's society, where transaction fraud poses an escalating threat, the development of robust Payment Security Systems is paramount, . This research is driven by the critical need to combat this evolving challenge. Preventing fraud demands not only vast amounts of data but also sophisticated models capable of accurate predictions. Our study rigorously compared several cutting-edge models, leveraging the power of SMOTE to further enhance their performance.

By comparing the performance metrics and visualizations of various models, we found that the Ensemble Model excels in multiple aspects. Specifically, the Ensemble Model achieved a precision of 0.98, indicating superior performance. This means that among all samples predicted as positive, the Ensemble Model has the highest correct prediction rate and the lowest false positive rate. Additionally, the AUC value of the Ensemble Model is 0.99, the highest among all models, demonstrating its ignificant aptitude to distinguish between positive and negative samples..

Although the Ensemble Model’s recall rate is 0.90, which is not the highest, it is still comparable to other models and maintains a high overall level. In terms of F1 Score, the Ensemble Model scored 0.94, consistent with Logistic Regression, XGBoost, LightGBM, CatBoost, and Neural Network, indicating a good balance between precision and recall.

Ensemble models leverage the predictions of multiple base models (such as LightGBM, XGBoost, and Neural Networks) to utilize the strengths of each model and compensate for the weaknesses of individual models\cite{lan2024improved, 10233897}. By incorporating the results of multiple models, ensemble models are more resilient to overfitting and underfitting issues associated with single models, resulting in a more stable overall model. By combining several models, ensemble models can better capture the diversity in data distribution, enhancing their generalization ability to new data. Additionally, when ensemble models are combined with relevant SMOTE techniques, their overall performance is significantly enhanced, maintaining a leading performance advantage. Our experiments have fully demonstrated the superior performance of ensemble models augmented with SMOTE techniques in fraud prediction.

\bibliographystyle{plain}
\bibliography{ref}

\end{document}